\documentclass[12pt]{iopart}
\usepackage{iopams}
\usepackage{graphicx}
\begin{document}

\title[]{Hydrodynamic modes, Green-Kubo relations,
and velocity correlations in dilute granular gases}

\author{J. Javier Brey \dag\, M. J. Ruiz-Montero \ddag, P. Maynar,
and M. I. Garc\'{\i}a de Soria}

\address{F\'{\i}sica Te\'{o}rica, Universidad de Sevilla, Apartado de Correos
1065, E-41080, Sevilla, Spain}

\ead{\dag\ brey@us.es, \ddag\ majose@us.es}

\begin{abstract}

It is shown that the hydrodynamic modes of a dilute granular gas
of inelastic hard spheres can be identified, and calculated in the
long wavelength limit. Assuming they dominate at long times,
formal expressions for the Navier-Stokes transport coefficients
are derived. They can be expressed in a form that generalizes the
Green-Kubo relations for molecular systems, and it is shown that
they can also be evaluated by means of $N$-particle simulation
methods. The form of the hydrodynamic modes to zeroth order in the
gradients is used to detect the presence of inherent velocity
correlations in the homogeneous cooling state, even in the low
density limit. They manifest themselves in the fluctuations of the
total energy of the system. The theoretical predictions are shown
to be in agreement with molecular dynamics simulations. Relevant
related questions deserving further attention are pointed out.

\end{abstract}

\submitto{\JPCM}


\maketitle

\section{Introduction}
\label{s1} Rapid granular flows exhibit a phenomenological
behavior that is similar in many ways to that of ordinary fluids.
Then it is not surprising that the methods of statistical
mechanics and kinetic theory are being successfully extended to
describe them. The primary feature of grains that must be captured
in any idealized model is the inelasticity of collisions. The
prototype is a system of smooth inelastic hard spheres (or disks
in two dimensions) whose energy loss in collisions is
characterized by a coefficient of normal restitution. In real
granular media, this coefficient depends on the relative velocity
of the colliding particles \cite{ByP03}, but for the sake of
simplicity, and expecting to retain the main qualitative features,
it will be considered here as a material constant.

At the most fundamental level of particle dynamics, the effect of
the inelasticity is easily incorporated by changing the binary
collision rule. As in the elastic case, the dynamics of the system
consists of  a sequence of free streaming between collisions and
instantaneous velocity changes on binary collisions. Therefore, a
statistical mechanics description can be formulated by means of
the Liouville equation of the system \cite{BDyS97,vNEyB98}. From a
conceptual point view, this description has the same degree of
justification as that of inherently non-equilibrium molecular
systems. Proceeding to a more mesoscopic description, as provided
by kinetic theory, is a more difficult and controversial problem.
Formally, the same methods as developed for hard sphere gases with
elastic collisions can be applied. In particular, in the low
density limit, the inelastic Boltzmann equation is derived. Thus,
it seems that the basis for this equation is the same as in the
elastic case. A different question is the domain in the parameter
space of density and coefficient of restitution in which it
actually provides an accurate description of the system
\cite{ByR04}.

The Boltzmann equation is a kinetic equation describing the time
evolution of the one-particle distribution function, that is an
average quantity. It is also possible to study,again in low
density limit, fluctuations and correlations in the positions and
velocities of the particles by means of the associated
distributions functions. It is known that correlations can become
quite large in out of equilibrium elastic systems, on the
appropriate long time and length scales. This should not be
interpreted as a signature of the failure of the Boltzmann
equation. The latter only requires that the precollisonal part of
the correlations of two particles at contact be small
\cite{KyS82,Lu96}. Closed kinetic equations for the fluctuation
and correlation functions valid in the low density limit, can be
derived inside the same approximation schemes that lead to the
Boltzmann equation. This includes formal expansions in a small
parameter \cite{Gr58} and cluster expansions of the reduced
distribution functions \cite{EyC81}. Both methods have been
considered for granular gases \cite{Du01,BGMyR04}. Nevertheless,
while the inelastic Boltzmann equation and its implications have
been extensively studied, little is known about fluctuations and
correlations in granular gases, even though it is widely
recognized that they are crucial for the understanding of the
properties of these systems.

As mentioned above, granular gases display fluid dynamical
behavior. Then, they are expected to obey some kind of
hydrodynamic equations, understood as a set of closed equations
for the hydrodynamics fields: density, flow velocity, and
(granular) temperature. In fact, more or less phenomenological
hydrodynamic equations have been successfully applied to many
different situations. An appropriate context in which the issue of
the existence of hydrodynamics for dilute granular gases, the form
of the hydrodynamic equations, and their range of validity, can be
addressed is provided by the inelastic Boltzmann equation. By
using an extension of the Chapmann-Enskog procedure, the
Navier-Stokes equations have been derived, and formal expressions
for the transport coefficients obtained. They are, in principle,
valid for arbitrary physical values of the coefficient of
restitution \cite{BDKyS98}. Recently, these expressions have been
expressed in the form of time integrals of the correlations of
some velocity functions computed in a given reference state, the
so-called homogeneous cooling state \cite{DyB02,BDyR03}.
Consistency requires that these integrals exist in the long time
limit. The above expressions are the generalization of the
Green-Kubo relations for elastic gases, although with
significative differences that could have been hardly guessed a
priori.

It must be stressed that the Chapmann-Enskog method and the
Green-Kubo relations presume the existence of hydrodynamics on the
appropriate length and time scales and, therefore, they do not add
any fundamental support to the validity of a hydrodynamic
description for granular gases. On the other hand, if
Navier-Stokes equations exist, the Chapmann-Enskog results provide
their explicit form. A deeper physical understanding of the
structure of the Green-Kubo relations and their validity follows
from the analysis of the linearized Boltzmann equation, and more
specifically of the spectrum of the associated operator
\cite{BDyR03}. It is possible to identify the hydrodynamic part of
the spectrum, that is the one determining the form of the
Green-Kubo relations. It is given by $d+2$ points that in the
elastic limit correspond to the $d+2$ collisional invariants,
where $d$ is the dimension of the system. The key remaining point
for the existence of hydrodynamics is whether the corresponding
modes dominate for long times and long wavelengths
\cite{DyB03,DyB04}. Quite interestingly, the hydrodynamic modes
also determine the form of the fluctuations of the hydrodynamic
quantities in granular systems, reflecting a generalization of the
fluctuation-dissipation theorems.

The aim of this paper is to review some recent advances and
present some new results along the above ideas. Particular
emphasis will be put on the relevance of the hydrodynamic modes
for the study of many fundamental and applied issues in granular
fluids, as it is the case for molecular, elastic systems. Along
the paper, and specially in the last Section, some of the many
remaining open questions in the context of the matter discussed
here will be indicated.

\section{The homogeneous cooling state of a dilute granular gas}
\label{s2} The system considered is composed of $N$ smooth
inelastic hard spheres ($d=3$) or disks ($d=2$) of mass $m$ and
diameter $\sigma$. Their dynamics consists of free streaming until
two particles $i$ and $j$ are at contact, when their velocities
change according with the rule
\begin{eqnarray}
\label{2.1} {\bi v}_{i} \rightarrow {\bi v}^{\prime}_{i} \equiv
b_{\bsigma} {\bi v}_{i} & = & {\bi v}_{i} -\frac{1+\alpha}{2}
\left( \widehat{\bsigma} \cdot {\bi v}_{ij} \right)
\widehat{\bsigma}, \nonumber \\
{\bi v}_{j} \rightarrow {\bi v}^{\prime}_{j} \equiv b_{\bsigma}
{\bi v}_{j} & = & {\bi v}_{j} +\frac{1+\alpha}{2} \left(
\widehat{\bsigma} \cdot {\bi v}_{ij} \right) \widehat{\bsigma},
\end{eqnarray}
where ${\bi v}_{ij}={\bi v}_{i}-{\bi v}_{j}$ is the relative
velocity, $\widehat{\bsigma}$ is the unit vector pointing from the
center of particle $j$ to the center of particle $i$ at contact,
and $\alpha$ is the coefficient of normal restitution. It is
defined in the range $0 < \alpha \leq 1$ and measures the
inelasticity of collisions. Here it will be taken as a constant.

>From the above dynamics, the Liouville equation governing the time
evolution of the $N$-particle distribution function of the system
can be derived \cite{BDyS97,vNEyB98}. An equivalent representation
is given by the BBGKY hierarchy for the $s$-particle distribution
functions ($s=1,2, \ldots$). In the low density limit, a closed
equation for the one-particle distribution function,
$f(\bi{r},\bi{v},t)$, can be derived by using the same procedures
as for elastic hard particles \cite{BDyS97,vNEyB98}. This is the
inelastic Boltzmann equation,
\begin{equation}
\label{2.2} \left( \frac{\partial}{\partial t} + {\bi v} \cdot
\frac{\partial}{\partial {\bi r}} \right) f({\bi r},{\bi v},t) =
J[f,f],
\end{equation}
where $J[f|f]$ is the inelastic Boltzmann collision operator
\cite{BDyS97,vNEyB98,LSJyCh84,JyR85,GySh95}.

For an isolated granular fluid, no homogenous steady solution is
possible. It is trivially seen that the energy of the system
decreases monotonically due to the inelasticity of collisions. On
the other hand, there is a spacially homogeneous state, the
so-called homogeneous cooling state (HCS), for which all the time
dependence occurs only through the granular temperature. The
latter is defined in the usual way from the average kinetic
energy. For the one-particle distribution, this implies a scaling
of the form \cite{GySh95}
\begin{equation}
\label{2.3} \fl f_{hcs}({\bi v},t)=n_{H} v_{0}^{-d}[T_{hcs}(t)]
\chi_{hcs} (c), \quad \quad  v_{0}[T_{hcs}(t)]=
\left[\frac{2k_{B}T_{hcs}(t)}{m} \right]^{1/2}.
\end{equation}
Here $n_{H}$ is the number of particles density, $k_{B}$ is the
Boltzmann constant, and $T_{hcs}(t)$ is the time-dependent
temperature of the HCS. Finally, $\chi_{hcs}(c)$ is an isotropic
function of ${\bi c}= {\bi v}/v_{0}[T(t)]$. Use of Eq.\
(\ref{2.3}) into Eq.\ (\ref{2.2}) provides an integral equation
for $\chi_{hcs}$,
\begin{equation}
\label{2.5} \frac{\zeta_{0}}{2} \frac{\partial}{\partial {\bi c}}
\cdot \left( {\bi c}\chi_{hcs}\right)=
J_{c}[\chi_{hcs},\chi_{hcs}],
\end{equation}
where $J_{c}$ is the scaled Boltzmann collision operator,
\begin{equation}
\label{2.6}  J_{c}[\chi_{hcs},\chi_{hcs}] = \int d{\bi c}_{1}
\overline{T}_{0}({\bi c},{\bi c}_{1}) \chi_{hcs}(c)
\chi_{hcs}(c_{1}),
\end{equation}
\begin{equation}
\label{2.6a} \overline{T}_{0}({\bi c},{\bi c}_{1})= \int d
\widehat{\bsigma}\, \Theta [{\bi g} \cdot \widehat{\bsigma}] {\bi
g} \cdot \widehat{\bsigma} \left[ \alpha^{-2}
b_{\bsigma}^{-1}({\bi c},{\bi c}_{1}) -1 \right],
\end{equation}
with ${\bi g} \equiv {\bi c}-{\bi c}_{1}$, $\Theta$ the Heaviside
step function, and $b_{\bsigma}^{-1}$ an operator replacing the
velocities ${\bi c}$ and ${\bi c}_{1}$ to its right by the
precollisional values obtained by inverting the collision rules
given in Eq.\ (\ref{2.1}). Moreover, the scaling property
(\ref{2.3}) implies that the time dependence of the temperature is
given by the Haff law
\begin{equation}
\label{2.7}
 T_{hcs}(t)=T_{hcs}(0) \left[ 1+\frac{1}{2}
\zeta_{hcs}(0) t \right]^{-2}.
\end{equation}
The cooling rate of the HCS, $\zeta_{hcs}$, is given by
\begin{equation}
\label{2.8} \zeta_{hcs}(t)=\frac{(1-\alpha^{2})\pi^{\frac{d-1}{2}}
\sigma^{d-1}n_{H}v_{0}(t)}{2\,  \Gamma \left( \frac{d+3}{2}
\right)d} \int d{\bi c} \int d{\bi c}_{1}\, g^{3}\chi_{hcs}(c)
\chi_{hcs}(c_{1}),
\end{equation}
and
\begin{equation}
\label{2.9} \zeta_{0}=\frac{\ell
\zeta_{hcs}(t)}{v_{0}[T_{hcs}(t)]},
\end{equation}
where $\ell = 1/n_{H} \sigma^{d-1}$ is proportional to the mean
free path of the particles. Since $\zeta_{hcs} \propto
T_{hcs}^{1/2}$, Eq.\ (\ref{2.7}) becomes independent of the
initial condition in the long time limit,
\begin{equation}
\label{2.10} T_{hcs} \sim 4(\bar{\zeta}t)^{-2}, \quad \quad
\bar{\zeta}=\frac{\zeta_{hcs}(t)}{T_{hcs}^{1/2}(t)}\, ,
\end{equation}
indicating that all the homogeneous cooling states of a given
system tend to converge in the long time limit \cite{BRyM04}. This
seems to suggest the existence of some kind of $H$ theorem for the
homogeneous Boltzmann equation for inelastic hard particles,
although no proof of such a theorem has been given up to date.

\section{Linearized Boltzmann equation and hydrodynamic modes}
\label{s3} Now, the initial value problem for small perturbations
around the HCS of an isolated granular gas will be considered. In
this case, the Boltzmann equation can be linearized around the
HCS, and it will be shown that, in appropriate dimensionless
units, its general solution can be related to a linear eigenvalue
problem. By considering times over which the deviation of the
distribution function from $f_{hcs}$ remains small, we can write
\begin{equation}
\label{3.1} f({\bi r},{\bf v},t)=f_{hcs}({\bi v},t)+\delta f({\bi
r},{\bf v},t), \quad \quad |\delta f({\bi r},{\bi v},t)| \ll
f_{hcs}({\bi v},t).
\end{equation}
In order to eliminate the time dependence of the HCS, it is
convenient to introduce the scaled velocity ${\bi c}$ defined
below  Eq.\ (\ref{2.3}) and the dimensionless quantities
\begin{equation}
\label{3.2} {\bi l}=\frac{\bi r}{\ell}, \quad s=\int_{0}^{t}
dt^{\prime}\, \frac{v_{0}(t^{\prime})}{\ell}\, , \quad \delta \chi
({\bi l},{\bi c},s)=n_{H}^{-1} v_{0}^{d}(t) \delta f ({\bi r},{\bi
v},t).
\end{equation}
The time scale $s$ is proportional to the average of the
accumulated number of collisions per particle in the interval
$(0,t)$.  Then, substitution of Eq.\ (\ref{3.1}) into the
Boltzmann equation and retaining only terms up to first order in
the deviations, gives the linear Boltzmann equation
\begin{equation}
\label{3.3} \frac{\partial}{\partial s}\delta \chi +{\bi c}\cdot
\frac{\partial}{\partial {\bi l}} \delta \chi= \Lambda \delta
\chi({\bi l},{\bi c},s),
\end{equation}
where the linear operator $\Lambda$ is given by
\begin{equation}
\label{3.4} \Lambda \delta \chi =\textit{J}_{c}[\chi_{hcs},\delta
\chi]+\textit{J}_{c}[\delta
\chi,\chi_{\mathrm{hcs}}]-\frac{\zeta_{0}}{2}\frac{\partial}{\partial
{\bi c}} \cdot \left( {\bi c} \delta \chi \right).
\end{equation}
In this representation, the effect of the cooling appears
explicitly through the last term on the right hand side of the
above expression.

Solutions to Eq.\ (\ref{3.3}) will be sought in a Hilbert space
defined by the scalar product
\begin{equation}
\label{3.5} <g|h>=\int d{\bf c}\, \chi_{hcs}^{-1}(c) g^{*}({\bf
c})h({\bf c}),
\end{equation}
where $g^{*}$ denotes the complex conjugate of $g$. It must be
noted that this is a stronger requirement than the necessary
normalization of the distribution function. Since the collision
operator $\Lambda$ does not affect the space variable, it is
enough to consider a single Fourier mode, i.e.,
\begin{equation}
\label{3.6} \delta \chi ({\bi l},{\bi c},s)= e^{i {\bi k} \cdot
{\bi l}} \delta \tilde{\chi} ({\bi k}, {\bi c},s),
\end{equation}
so that the linearized Boltzmann equation becomes
\begin{equation}
\label{3.7} \frac{\partial}{\partial s}\ \delta \tilde{\chi}
=\left( \Lambda - i{\bi k} \cdot {\bi c} \right) \delta
\tilde{\chi}.
\end{equation}
All linear excitations are determined from the spectrum of the
operator $\Lambda - i {\bi k} \cdot {\bi c}$. The hydrodynamic
part of this spectrum is defined as follows. For $k=0$ it is given
by those eigenvalues that coincide with the eigenvalues of the
balance equations for the momentum density, flow velocity, and
temperature obtained from the linearized homogeneous Boltzmann
equation. Assuming that the cooling rate for homogeneous systems
can be expressed as a function of the density and the temperature,
these eigenvalues are found to be \cite{BDyR03}
\begin{equation}
\label{3.8} \lambda_{1}=0, \quad \lambda_{2}=\zeta_{0}/2\,  \quad
\lambda_{3}=-\zeta_{0}/2,
\end{equation}
the eigenvalue $\lambda_{2}$ being $d$-fold degenerated. For
finite $k$, the hydrodynamic modes are defined as the solutions of
the equation
\begin{equation}
\label{3.9} \left( \Lambda - i {\bi k} \cdot {\bi c}
\right)\xi_{i}({\bi k},{\bf c})=\lambda_{i} \xi_{i}({\bi k},{\bf
c})
\end{equation}
that are continuously connected as functions of ${\bi k}$ to the
above special solutions for $k=0$. Of course, consistency requires
that $\Lambda$ includes the spectrum of the balance equations for
$k=0$. Straightforward calculations show that \cite{BDyR03}
\begin{equation}
\label{3.10} \Lambda \xi_{i}({\bi c})=\lambda_{i} \xi_{i}({\bi
c}), \quad i=1,2,3
\end{equation}
with the $\lambda_{i}$'s given by Eq.\ (\ref{3.8}) and
\begin{equation*}
\xi_{1}({\bi c})= \chi_{hcs}(c)+\frac{\partial}{\partial {\bi c}}
\cdot \left[ {\bi c} \chi_{hcs} (c) \right],
\end{equation*}
\begin{equation}
\label{3.11} {\bxi}_{2}({\bi c})=-\frac{\partial
\chi_{hcs}(c)}{\partial {\bi c}}, \quad \xi_{3}({\bi c})=
-\frac{\partial}{\partial {\bi c}} \cdot \left[ {\bi c} \chi_{hcs}
(c) \right].
\end{equation}
This confirms the existence of the hydrodynamic modes. Their
evaluation for finite $k$ is now a technical problem. To
Navier-Stokes order, they have been computed by using perturbation
expansion methods \cite{DyB03}, and the results agree with those
obtained from the Navier-Stokes equations derived by means of the
Chapmann-Enskog procedure \cite{BDKyS98}. Here an alternative
approach, designed to derive in a direct way expressions for the
Navier-Stokes transport coefficients, will be outlined.

As a consequence of $\Lambda$ being not Hermitian, the
eigenfunctions $\xi_{i}$ are not orthogonal. It is then useful to
introduce a set of functions $\{\bar{\xi}_{i} \}$ being
biorthogonal to the set $\{ \xi_{i}, i=1,2,3 \}$. A convenient
choice is
\begin{equation}
\label{3.12}  \{ \overline{\xi}_{i} \}= \left\{ \chi_{hcs}(c),{\bi
c} \chi_{hcs}(c),\left( \frac{c^{2}}{d} +\frac{1}{2} \right)
\chi_{hcs}(c) \right\}, \quad   \langle \overline{\xi}_{i}
|\xi_{j} \rangle =\delta_{i,j}\, .
\end{equation}

The general solution to the linearized Boltzmann equation in the
Hilbert space can be formally written as
\begin{equation}
\label{3.13} \delta \tilde{\chi} ({\bi k}, {\bi c},s)=
\sum_{i}\tilde{a}_{i}({\bi k},s) \chi_{i} ({\bi k},{\bi c})+
\delta \tilde{\chi}^{m} ({\bi k}, {\bi c},s),
\end{equation}
where the sum extends over the $d+2$ hydrodynamic modes and the
term $\delta \tilde{\chi}^{m}$ contains all the other
``microscopic'' excitations. {\it Assuming} that the hydrodynamic
spectrum dominates for long times and small gradients, $\delta
\tilde{\chi}^{m}$ can be neglected, and the coefficients
$\tilde{a}_{i}$ are given by
\begin{equation}
\label{3.14} \tilde{a}_{i}({\bi k},s)=\langle \bar{\xi}_{i}|
\delta \tilde{\chi} ( {\bi k},s) \rangle.
\end{equation}
These coefficients are combinations of the hydrodynamic fields,
i.e. they can be expressed in terms of the deviations of the
macroscopic density $n$, temperature $T$ and fluid velocity ${\bi
u}$. It is now possible to derive a formal expression for  $\delta
\tilde{\chi} ({\bi k}, {\bi c},s)$ to first order in $k$, and from
it the Navier-Stokes order for the pressure tensor and the heat
conductivity \cite{BDyR03}. The results will be given in the next
section.

The above discussion relies on the assumption that the
hydrodynamic point spectrum found here is isolated and smaller
than all the other parts of the spectrum, so that it dominates for
long times and long wavelengths. This is a necessary condition for
the existence of hydrodynamics and has been proven in the elastic
case. The proof does not extend directly to the inelastic case,
although there are some reasons to expect that this is the case
even for rather strong inelasticity. A detailed discussion of the
basis for a hydrodynamic description of a granular gas in this
context is given in \cite{DyB04}.

\section{Green-Kubo representation of the transport coefficients}
\label{s4} The procedure described in the previous Section leads
to the following expressions for the dissipative part of the
pressure tensor $\Pi_{ij}({\bi r},t)$ and the heat flux ${\bi
q}({\bi r},t)$ to Navier-Stokes order\cite{BDyR03}:
\begin{equation}
\label{4.1} \Pi_{ij}= - \eta \left( \frac{\partial u_{i}}{\partial
r_{j}}+\frac{\partial u_{j}}{\partial r_{i}} -\frac{2}{d}
\delta_{ij} \bnabla \cdot {\bf u} \right), \quad {\bi q}= -\kappa
\bnabla T -\mu \bnabla n,
\end{equation}
where $\eta$ is the shear viscosity, $\kappa$ the (thermal) heat
conductivity, and $\mu$ a new transport coefficient that vanishes
in the elastic limit and that will be referred to as the diffusive
heat conductivity. They are given by
\begin{equation}
\label{4.2} \eta = n_{H} m \ell v_{0}(T) \int_{0}^{\infty} ds\,
e^{-s \zeta_{0}/2} \langle \Delta_{xy}(s) \Phi_{2,xy} \rangle ,
\end{equation}
\begin{equation}
\label{4.3} \kappa=n_{H} \ell v_{0}(T) \int_{0}^{\infty} ds\, e^{s
\zeta_{0}/2} \langle \Sigma_{x}(s) \Phi_{3,x} \rangle ,
\end{equation}
\begin{equation}
\label{4.4} \mu= m \ell v_{0}^{3}(T) \left[ \int_{0}^{\infty} ds
\left( e^{s \zeta_{0}/2} - 1 \right) \langle \Sigma_{x}(s)
\Phi_{3,x} \rangle +\frac{1}{2} \int_{0}^{\infty} ds\, \langle
\Sigma_{x}(s) c_{x} \rangle \right].
\end{equation}
In the above expressions, $\Delta_{xy}$ and $\Sigma_{x}$ are the
transversal momentum flux and the heat flux, respectively,
\begin{equation}
\label{4.5} \Delta_{xy}({\bi c})= c_{x}c_{y}, \quad \quad
\Sigma_{x}({\bi c})=\left( c^{2}-\frac{d+2}{2} \right)c_{x},
\end{equation}
while $\Phi_{2,ij}$ and $\Phi_{3,x}$ are ``modified fluxes'',
\begin{equation}
\label{4.6} \Phi_{2,xy}({\bi c})= -c_{x} \frac{\partial \ln
\chi_{hcs} (c)}{\partial c_{y}}, \quad \Phi_{3,x}({\bi c})=
-\frac{c_{x}}{2} \left[ d+ {\bi c} \cdot \frac{\partial  \ln
\chi_{hcs}(c) )}{\partial {\bi c}} \right].
\end{equation}
The angular brackets denote ``time-correlation functions'' in the
HCS,
\begin{equation}
\label{4.7} \langle g(s) h \rangle \equiv \int d{\bi c}\,
\chi_{hcs}(c) g({\bi c},s) h({\bi c})
\end{equation}
where the time dependence is given by
\begin{equation}
\label{4.8} g({\bi c},s)= e^{s \tilde{\Lambda}({\bi c})} g({\bi
c}),
\end{equation}
with
\begin{equation}
\label{4.9} \fl \tilde{\Lambda}({\bf c})= \int d {\bi c}_{1}\,
\chi_{hcs} (c_{1}) \int d\widehat{\bsigma}\, \Theta ({\bi g}\cdot
\widehat{\bi \sigma}) {\bf g}\cdot \widehat{
\bsigma}[b_{\bsigma}({\bi c},{\bi c}_{1})-1] (1+P)+
\frac{\zeta_{0}}{2} {\bi c} \cdot \frac{\partial}{\partial {\bi
c}}\, .
\end{equation}
Here, $P$ is an operator that interchanges the velocities ${\bi
c}$ and ${\bi c}_{1}$ to its right. Equations
(\ref{4.2})-(\ref{4.4}) are the generalization for inelastic hard
collisions of the dilute limit of the well-known Green-Kubo
relations for molecular fluids, reducing to them in the elastic
limit \cite{DyB02}. They can be understood as generalized
fluctuation-dissipation relations, in the sense that they relate
the transport coefficients characterizing dissipation in
inhomogeneous systems with the time decay of fluctuations in the
homogeneous reference state (the HCS). Equivalent expressions for
the transport coefficients are obtained by solving the nonlinear
Boltzmann equation by the Chapman-Enskog method
\cite{BDKyS98,DyB02} and also by perturbation analysis of the
hydrodynamic spectrum of the linear Boltzmann equation
\cite{DyB03}. The effect of the inelasticity of collisions
manifests itself in several ways. Perhaps the most unexpected
among them is that the initial dissipative fluxes $\Delta_{xy}$
and $\Sigma_{x}$ are replaced by the modified fluxes $\Phi_{2,xy}$
and $\Phi_{3,x}$ in the time correlation functions appearing in
the expressions of $\eta$ and $\kappa$. This is a direct
consequence of the form of the hydrodynamic modes. While in a
molecular system they are given by linear combinations of the
Maxwellian times $1$, ${\bi c}$, and $c^{2}$, in granular gases
they are given by Eq.\ (\ref{3.11}), and it should be realized
that $\chi_{hcs}(c)$ is not a Maxwellian outside the elastic limit
$\alpha=1$.

Since the diffusive heat conductivity $\mu$ is a peculiarity of
granular systems, it seems appropriate to comment about its
relevence. In general, its contribution to the heat flux is much
smaller than the one coming from the usual heat conductivity
associated with the temperature gradient, because $\kappa \gg \mu$
for all $\alpha$. Nevertheless, there are some cases where the
existence of this contribution manifests itself in a clear way.
Consider, for instance, an open granular system in presence of
gravity. The system is fluidized by means of a vibrating bottom
plate. In the steady state, the granular temperature profile
presents a minimum at a certain height, increasing from there on.
This temperature inversion has been observed in computer
simulations \cite{BRyM01,HByH97,RyS03} and also in experiments
\cite{CyR91,WHyP01,ByK03,HYCMyW04}. Its theoretical explanation is
tied to the existence of the diffusive contribution to the heat
flux \cite{BRyM01}, and the value of the transport coefficient
$\mu$ can be directly obtained from the measurement of the heat
flux at the temperature minimum \cite{ByR04a}. For a low density
gas, the results are in good agreement with the theoretical
predictions discussed in the next Section.

\section{Simulation study of the Green-Kubo relations}
\label{s5} The formal expressions of the transport coefficients
given by Eqs. (\ref{4.2})-(\ref{4.4}) are rather involved and have
been evaluated analytically only in some approximation, namely by
introducing an expansion in orthogonal functions, keeping only
rather arbitrarily the lowest orders contributions
\cite{BDKyS98,SyG98,ByC01}. On the other hand, the above
Green-Kubo relations can be transformed into a form that is
suitable for evaluation by means of $N$-particle simulation
techniques. Two main points must be addressed for that. First,
there is the technical difficulty that the operator
$\tilde{\Lambda}$ defining the time dependence of the fluxes,
involves the cooling rate $\zeta_{0}$ that, therefore, should be
known {\it a priori} and it is not. This can be overcome by making
a change in the time scale and realizing the existence of an exact
mapping of the HCS onto a steady state \cite{L01,BRyM04}. The
second issue is to transform Eqs. (\ref{4.2})-(\ref{4.4}) into
expressions involving a well-defined $N$-particle dynamics,
instead of the effective Boltzmann dynamics defined by
$\tilde{\Lambda}$. This can be done by using the same kind of
assumptions as needed to derive the Boltzmann equation and, in
particular, assuming that velocity correlations in the HCS are
negligible in the low density limit \cite{BRyM04}.

We have used the direct simulation Monte Carlo method (DSMC)
\cite{Bi94} to simulate the $N$-particle dynamics of a dilute
granular gas composed of inelastic hard spheres ($d=3$). The
details of the practical implementation of the above ideas and the
simulation technique are discussed in refs. \cite{BRyM04} and
\cite{ByR04b}. Here it will only be stressed that the DSMC allows
to simulate a homogeneous state in such a way that it stays
homogeneous ``by definition'', avoiding the spontaneous developing
of spacial inhomogeneities. This is important in the present
context, since the HCS is known to be unstable against spacial
long wave perturbations, even in the low density limit. This is
the so-called clustering instability \cite{GyZ93,BRyC96}.

\begin{figure}
\begin{center}
\includegraphics[scale=0.35,angle=-90]{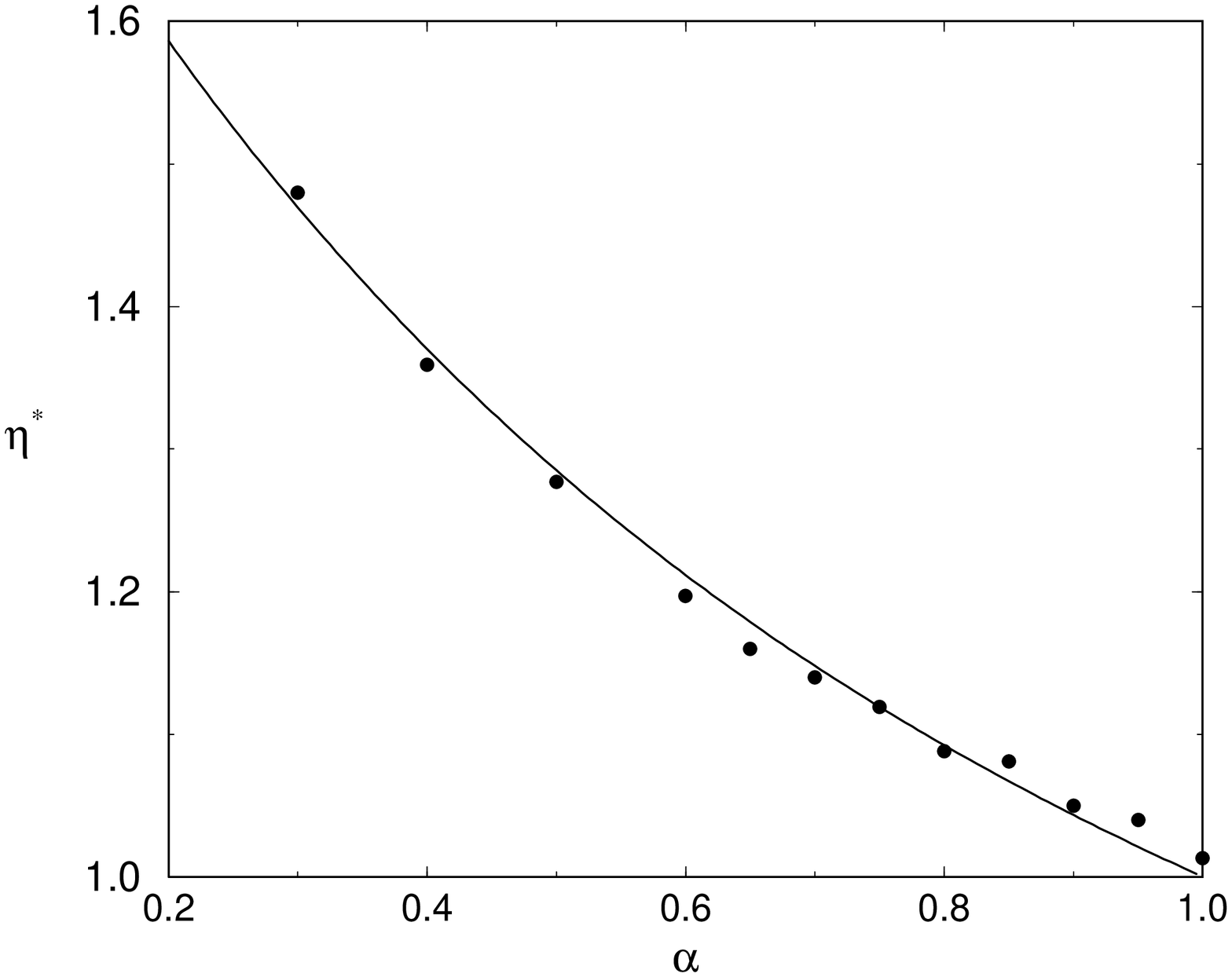} \quad
\includegraphics[scale=0.35,angle=-90]{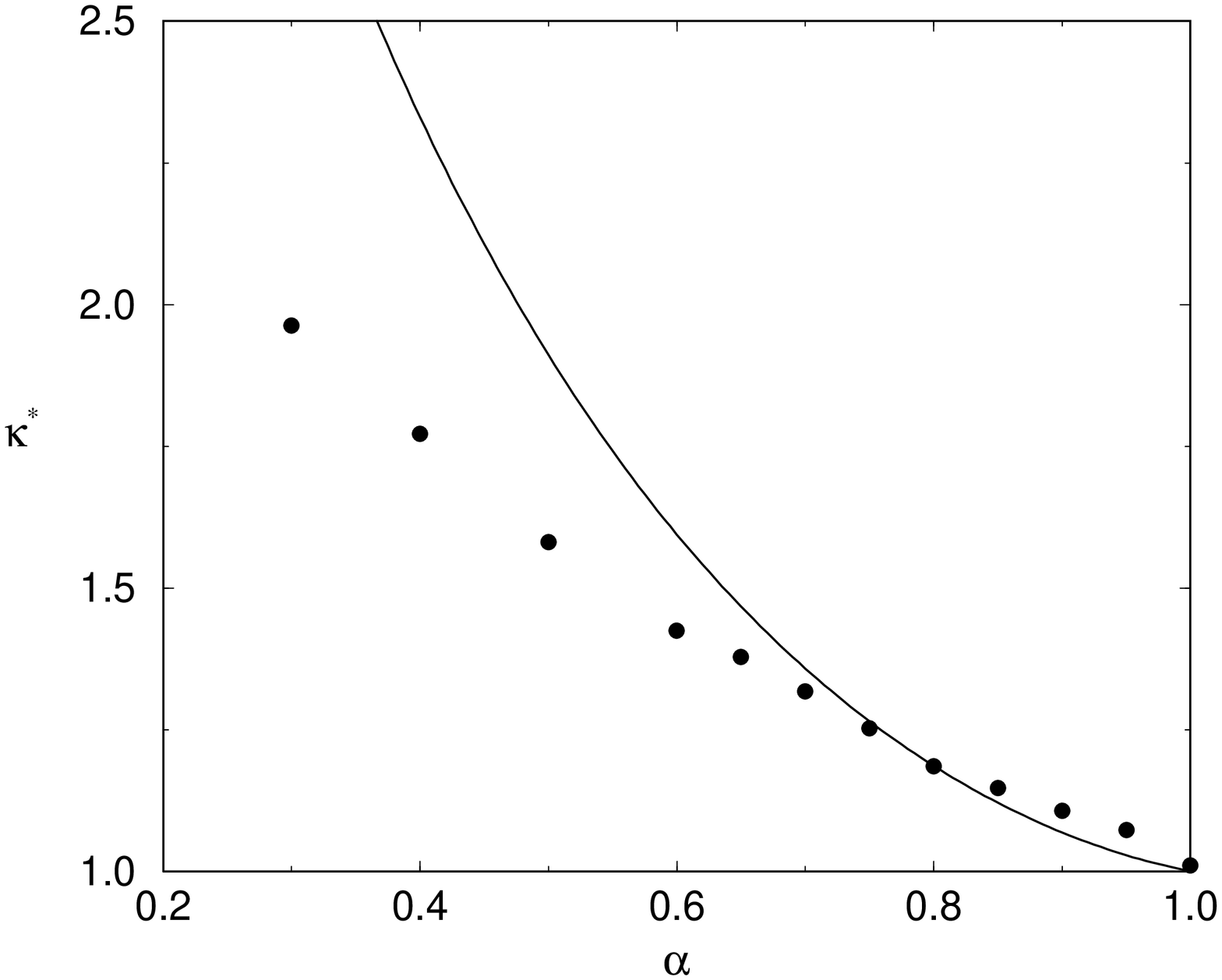}
\caption{Left: The dimensionless reduced coefficient of shear
viscosity $\eta^{*}$ as a function of $\alpha$. The symbols are
from the DSMC method, while the solid line is the analytical
result obtained in the first Sonine approximation \cite{BDKyS98}.
Right: The same for the reduced coefficient of thermal heat
conductivity $\kappa^{*}$ }\label{f1}
\end{center}
\end{figure}

In the simulations, it is observed that the HCS, or more precisely
its equivalent steady representation, is reached after a few
collisions per particle. This supports that the hydrodynamic point
spectrum identified in Sec.\ \ref{s4} is isolated and smaller than
all the other parts of the spectrum. Moreover,  all the time
correlation functions appearing in Eqs. (\ref{4.2})-(\ref{4.4})
are found to decay exponentially, at least until they decay by two
orders of magnitud from their initial values. This has been
exploited in order to compute the time integrals. The results
obtained for the shear viscosity are shown in Fig. \ref{f1}
(left). What is actually plotted is the shear viscosity scaled
with its elastic value in the first Sonine approximation, i.e.
\begin{equation}
\label{5.1} \eta^{*} \equiv \frac{\eta(T)}{\eta_{0}(T)}, \quad
\eta_{0}(T)=\frac{5 (mk_{B}T)^{1/2}}{16 \pi^{1/2} \sigma^{2}}\, .
\end{equation}
Also plotted in the same figure is the approximate theoretical
prediction obtained when computing Eq.\ (\ref{4.2}) in the first
Sonine approximation \cite{BDKyS98}. It is observed that the
viscosity monotonically increases with the inelasticity of
collisions (decreasing $\alpha$) and that the theoretical
prediction is in good agreement with the simulation results over
all the range of values of $\alpha$.

The simulation results for the (thermal) heat conductivity are
given in Fig. \ref{f1} (right). Similarly to above, the
dimensionless plotted quantity is
\begin{equation}
\kappa^{*} \equiv \frac{\kappa(T)}{\kappa_{0}(T)}, \quad
\kappa_{0}(T)=\frac{75 k_{B} (k_{B}T)^{1/2}}{64 (\pi
m)^{1/2}\sigma^{2}}.
\end{equation}
Again, it is found that the transport coefficient is a monotonic
decreasing function of $\alpha$. However, in this case the first
Sonine approximation leads to significant quantitative
discrepancies for small values of $\alpha$. This may be, at least
partially, due to the fact that the fluxes appearing in the
expression of the heat conductivity involve higher velocity
moments than those present in the expression of the shear
viscosity.

\begin{figure}
\begin{center}
\includegraphics[scale=0.35,angle=-90]{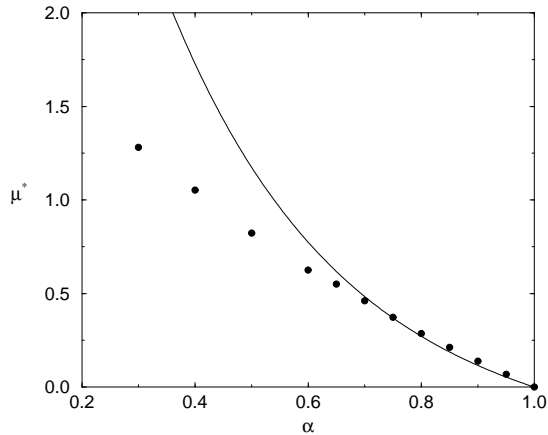}
\caption{The same as in Fig. \ref{f1}, but for the reduced
coefficient of diffusive heat conductivity $\mu^{*}$} \label{f2}
\end{center}
\end{figure}

The dimensionless reduced  diffusive heat conductivity $\mu^{*}$
has been defined as
\begin{equation}
\label{5.2} \mu^{*} \equiv \frac{ n \mu (T)}{T \kappa_{0}(T)},
\end{equation}
and the results obtained for it are presented in Fig.\ \ref{f2}.
As it is the case for the thermal heat conductivity, the first
Sonine approximation significantly overestimates this transport
coefficient for strong dissipation.

The analysis of the transport coefficients for a dilute gas of
inelastic hard disks ($d=2$) has been reported in \cite{ByR04b}.
The results are quite similar, the main difference being that the
discrepancies of the first Sonine approximation predictions are
stronger for $d=2$ than for $d=3$, as it is usually the case.

As already pointed out, in order to transform the effective
one-particle dynamics, implicit in the linearized evolution
operator $\tilde{\Lambda}$, into an $N$-particle dynamics, it has
been assumed that velocity correlations are negligible in the HCS
of a low density granular gas. Nevertheless, when the DSMC data
are carefully analyzed, it is seen that for strong inelasticity
($\alpha \lesssim 0.6$), velocity correlation effects are clearly
identified \cite{ByR04b}. For instance, their contribution to the
initial value of the time-correlation functions appearing in the
$N$-particle representation of the Green-Kubo expressions can be
up to $10 \%$ of the total value. This raises some fundamental
issues. Of course, a first possibility is that the velocity
correlations be just an artifact introduced by the simulation
method itself, i.e. an inaccuracy of the DSMC method. On the other
hand, if they are not, how can these correlations be incorporated
in a theoretical description of dilute granular gases? Is it
consistent with the description provided by the inelastic
Boltzmann equation? A partial answer to these questions will be
given in the next Section.

\section{Velocity correlations and total energy fluctuations in the HCS}
\label{s6} In spite of the great advance in the understanding of
rapid granular flows in the last years, little is known about
fluctuations and correlations in those systems. Most previous
studies have focussed on the study of the initial build-up of
spacial correlations in the development of the clustering
instability \cite{vNEByO97,BMyR98} or on the short time and length
scales precollisional velocity correlations when starting from a
thermal equilibrium state \cite{SPyM01}. Recently,  an empirical
expression for the two-particle distribution function of a system
in the HCS, based on simplicity and symmetry considerations, has
been proposed \cite{PByS02}. It includes both spatial and velocity
correlations, but their relation with the dynamics of the system
has not been analyzed.

Closed equations for the distribution functions characterizing
correlations in a dilute granular gas, can be derived by using the
same kind of approximations as needed to derive the inelastic
Boltzmann equation. In particular, in the hierarchy method it is
assumed that the order in the density of the precollisional part
of the correlation functions of two particles at contact,
increases as the number of involved particles increases
\cite{EyC81}. Of course, the results obtained under this
assumption must be consistent with it.

Here we will restrict ourselves to a dilute gas in the HCS. It is
assumed that the Liouville equations admits a solution describing
this state and having the scaling form
\begin{equation}
\label{6.1} \rho_{hcs}(\{ {\bi r}_{i},{\bi v}_{i}\},t) = [
v_{0}(t)]^{-dN} \rho^{*}_{hcs} \left( \{ {\bi r}_{ij} \} , \{ {\bi
c}_{i} \} \right),
\end{equation}
that is invariant under space translations. This implies an
analogous scaling for all the distribution functions of the system
and, in particular, leads to Eq.\ (\ref{2.3}) for the one-particle
distribution function. Two-particle correlations are described by
means of the function $g_{hcs}({\bi r}_{12}, {\bi v}_{1},{\bi
v}_{2},t)$, defined through the cluster expansion of the
two-particle distribution function $f_{2,hcs}({\bi r}_{12},{\bi
v}_{1},{\bi v}_{2},t)$,
\begin{equation}
\label{6.2} f_{2,hcs}({\bi r}_{12},{\bi v}_{1},{\bi
v}_{2},t)=f_{hcs}({\bi v}_{1},t)f_{hcs}({\bi
v}_{2},t)+g_{hcs}({\bi r}_{12}, {\bi v}_{1}, {\bi v}_{2},t).
\end{equation}
As a consequence of the scaling property (\ref{6.1}), the reduced
distribution function
\begin{equation}
\label{6.3} \tilde{g}_{hcs}({\bi l}_{12},{\bi c}_{1},{\bi c}_{2})=
n_{H}^{-1} \ell^{d} v_{0}^{2d}(t) g_{hcs}({\bi r}_{12},{\bi
v}_{1}, {\bi v}_{2},t)
\end{equation}
does not depend on $s$. Then, from the second equation of the
BBGKY hierarchy, neglecting the term involving three-particle
correlations, a closed  equation for $\tilde{g}_{hcs}$ is found
\cite{BGMyR04},
\begin{equation}
\label{6.4} \fl \left[ {\bi c}_{12} \cdot \frac{\partial}{\partial
{\bi l}_{12}} - \Lambda ({\bi c}_{1})-\Lambda ({\bi c}_{2})
\right] \tilde{g}_{hcs}({\bi l}_{12},{\bi c}_{1},{\bi c}_{2})  =
\delta ({\bi l}_{12}) \overline{T}_{0}({\bi c}_{1},{\bi c}_{2})
\chi_{hcs}(c_{1}) \chi_{hcs}(c_{2}).
\end{equation}
In order to focus on the velocity correlations, we introduce the
marginal velocity correlation function
\begin{equation}
\label{6.5} \psi_{hcs}({\bi c}_{1}, {\bi c}_{2}) \equiv \int d{\bi
l}_{12} \tilde{g}_{hcs}({\bi l}_{12},{\bi c}_{1};{\bi c}_{2}) +
\delta( {\bi c}_{1}-{\bi c}_{2}) \chi_{hcs}(c_{1}).
\end{equation}
Integration of Eq.\ (\ref{6.4}) yields
\begin{equation}
\label{6.6} \left[ \Lambda({\bi c}_{1})+\Lambda ({\bi c}_{2})
\right] \psi_{hcs}({\bi c}_{1},{\bi c}_{2})
=-\overline{T}_{0}({\bi c}_{1},{\bi c}_{2}) \chi_{hcs}(c_{1})
\chi_{hcs}(c_{2}).
\end{equation}

Trying to find the complete solution of the above equation is a
formidable task. We will restrict ourselves to find the
hydrodynamic part $\psi_{hcs}^{(h)}$ of the solution, defined by
\begin{equation}
\label{6.7} \psi_{hcs}^{(h)}({\bi c}_{1},{\bi c}_{2}) \equiv
\mathcal{P} \psi_{hcs}({\bi c}_{1},{\bi c}_{2}),
\end{equation}
where $\mathcal{P}$ is the projection operator onto the direct
product of the hydrodynamic part of Hilbert space. Next, we apply
$\mathcal{P}$ to both sides of Eq.\ (\ref{6.6}) and, in order to
get a closed equation for $\psi_{hcs}^{(h)}$, we make the
approximation $\mathcal{P} \Lambda ({\bi c}_{i}) = \mathcal{P}
\Lambda ({\bi c}_{i}) \mathcal{P}$. Then, the resulting equation
can be solved yielding \cite{BGMyR04}
\begin{equation}
\label{6.8} \psi_{hcs}^{(h)}({\bi c}_{1},{\bi
c}_{2})=a_{33}(\alpha) \xi_{3}({\bi c}_{1}) \xi_{3} ({\bi c}_{2}).
\end{equation}
The coefficient $a_{33}(\alpha)$ is a functional of $\chi_{hcs}$,
vanishing in the limit $\alpha \rightarrow 1$. An explicit
expression can be obtained in the first Sonine approximation
\cite{BGMyR04}. The fact that there are no contributions to
$\psi_{hcs}^{(h)}$ from the other combinations of the hydrodynamic
modes is a consequence of the symmetry of the HCS and the strict
conservation of the number of particles and the total momentum. On
the other hand, Eq.\ (\ref{6.8}) clearly shows the presence of
velocity correlations that are due to the inelasticity of
collision being, therefore, inherent to granular systems.

A direct test of the accuracy of the above expression by means of
$N$-particle simulation is beyond the present possibilities.
Nevertheless, it is possible to carry out an indirect check by
computing the fluctuations of the total energy $E$. They are given
by
\begin{eqnarray}
\label{6.9}  \langle [ \delta E(t)]^{2} \rangle _{hcs} &\equiv&
\langle E^{2}(t) \rangle_{hcs}- \langle E(t) \rangle_{hcs}^{2}
\nonumber \\
&= & N k_{B}^{2} T^{2}_{hcs}(t) \int d{\bi c}_{1} \int d{\bi
c}_{2} c_{1}^{2} c_{2}^{2} \psi_{hcs}^{(h)}({\bi c}_{1},{\bi
c}_{2})  \nonumber \\
&=& N k_{B}^{2} T_{hcs}^{2}(t) e(\alpha),
\end{eqnarray}
with $e(\alpha)=d^{2} a_{33}(\alpha)$. In the above expressions,
$\langle \cdots \rangle_{hcs}$ denotes ensemble average with
$\rho_{hcs}(\{ {\bi r}_{i},{\bi v}_{i}\},t)$.

\begin{figure}
\begin{center}
\includegraphics[scale=0.35,angle=-90]{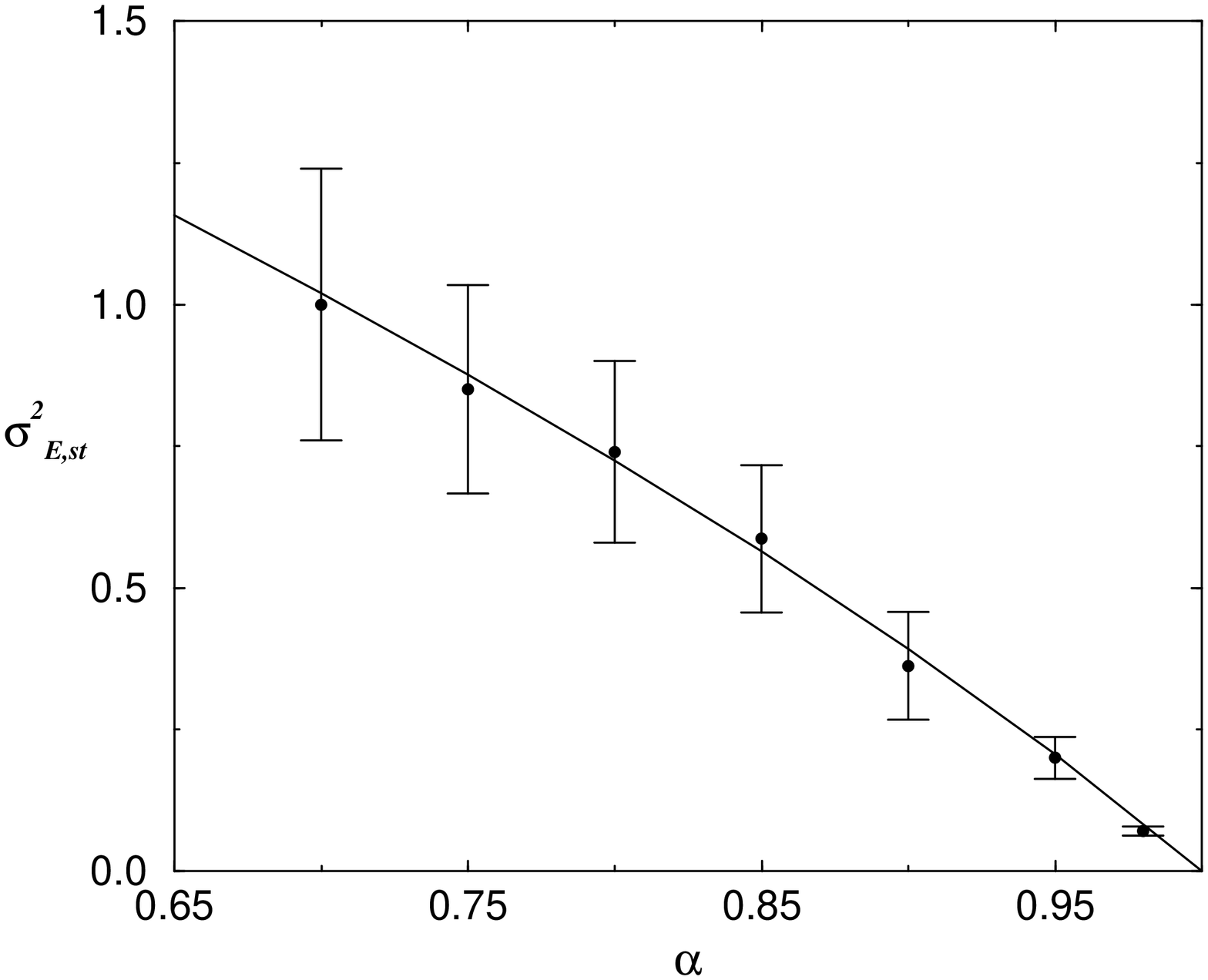} \quad
\includegraphics[scale=0.35,angle=-90]{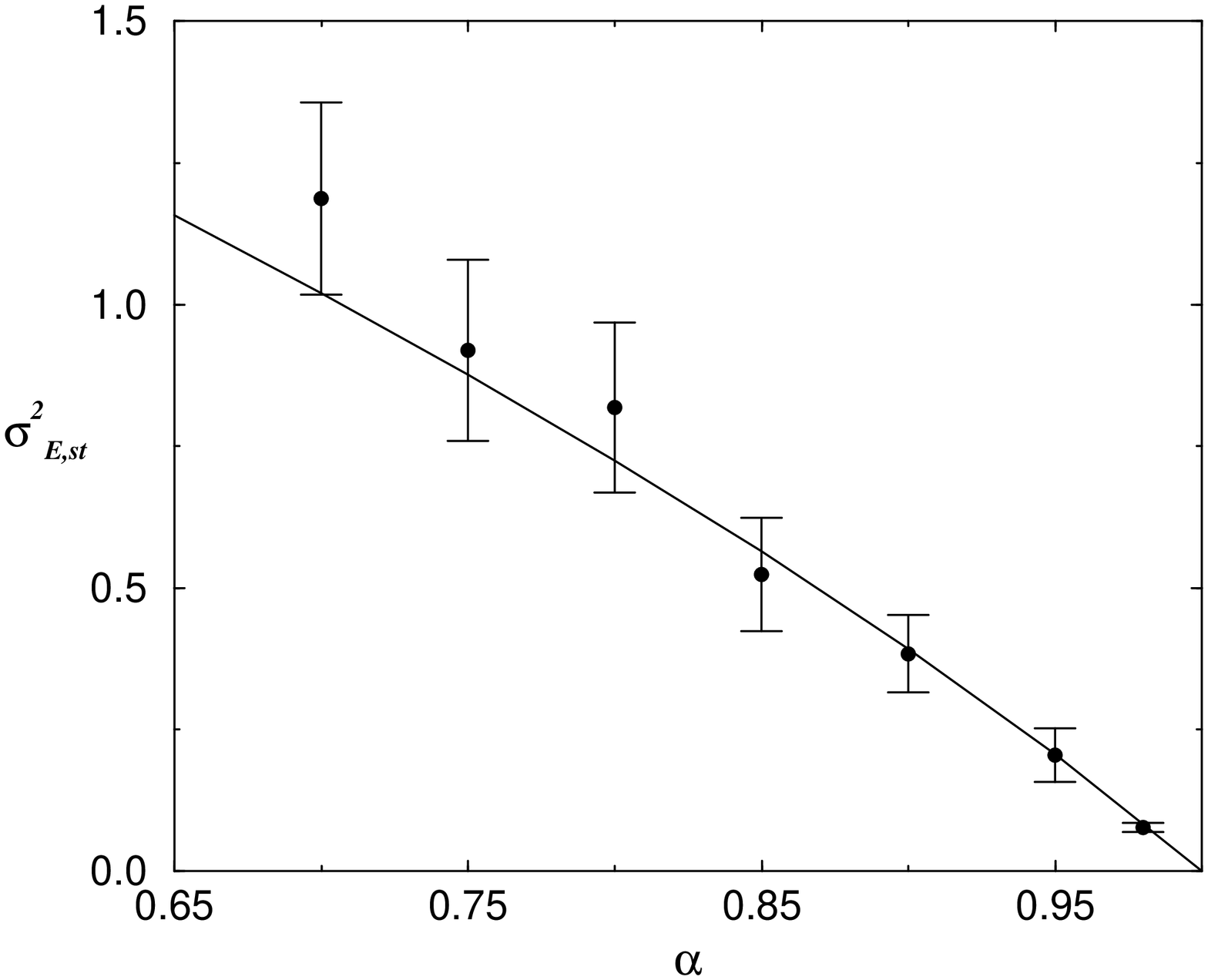}
\caption{Scaled second moment of the energy fluctuations
$\sigma_{E,st}^{2}$  as a function of the restitution coefficient
$\alpha$. The symbols are from MD simulations of a system of hard
disks and the solid line the theoretical prediction given by Eq.\
(\ref{6.9}). Left: $n_{H}=5 \times 10^{-3} \sigma^{-2}$;  right:
$n_{H}=0.01 \sigma^{-2}$ }\label{f3}
\end{center}
\end{figure}

To put the above expression in a proper context, it is important
to realize that, although its derivation relies on the knowledge
of the hydrodynamic modes, the validity of the result is not
restricted to any kind of hydrodynamic regime. On the contrary, it
is a general result for the HCS, assuming it exits, with the only
operational approximation given above Eq.\ (\ref{6.8}). We have
measured the energy fluctuations by performing Molecular Dynamics
simulations of a freely evolving system of inelastic hard disks.
Periodic boundary conditions were used, and the system size was
always chosen smaller than the critical value needed for the
spontaneous formation of spacial instabilities \cite{GyZ93}. The
results obtained are given in Fig. \ref{f3} for two small
densities, namely $n_{H}=5 \times 10^{-3} \sigma^{-2}$ and $n_{H}=
0.01 \sigma^{-2}$. The plotted quantity is $\sigma_{E,st}^{2}
\equiv \langle (\delta E)^{2}\rangle_{hcs}/ Nk_{B}^{2}
T_{hcs}^{2}$, which does not depend on time, and each of the
reported values corresponds to an average over several
trajectories. Also plotted is the theoretical prediction provided
by Eq.\ (\ref{6.9}), i.e. the function $e(\alpha)$ computed in the
first Sonine approximation. A quite good agreement is observed
over all the range of $\alpha$ considered, specially at the lowest
density. This  was expected given the low density character of the
theoretical approach sketched above. Unfortunately, considering
smaller values of the restitution coefficient requires the
simulations of  too small systems in order to avoid the
instability mentioned above, and consequently finite size effects
may become relevant.

Then, we conclude that there are intrinsic velocity correlations
in a granular fluid as a consequence of the inelasticity of
collisions, even in the low density limit. Moreover, they increase
as the value of $\alpha$ decreases, as expected. Also, the results
in Fig.\ \ref{f3} indicate that what we have termed hydrodynamic
part of the velocity correlations is well described by a
consistent low density theory, at least for $\alpha \gtrsim 0.7$.
On the other hand, it is not possible to infer anything more
precise from the results in this Section with regard to the
manifestation of velocity correlation effects in the evaluation of
the Green-Kubo expressions, as discussed in the previous Section.
There are two reasons for this: 1) it is easily realized that the
velocity correlations involved in the Green-Kubo expressions lie
outside the hydrodynamic subspace, and 2) as already indicated,
their effect become appreciable for $\alpha \lesssim 0.6$, i.e.
for systems more inelastic than those consider in the MD
simulations reported here.

\begin{figure}
\begin{center}
\includegraphics[scale=0.35,angle=-90]{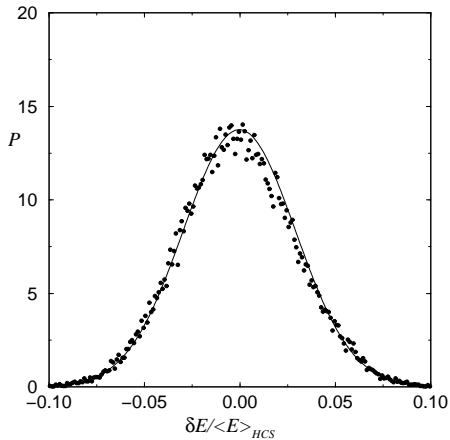}
\caption{Normalized distribution of the relative energy
fluctuations for a system of inelastic hard disks with $n_{H}=0.02
\sigma^{-2}$. The symbols are from MD simulations and the solid
line is a fit to a Gaussian.} \label{f4}
\end{center}
\end{figure}

>From the MD simulation results, it is also possible to compute the
normalized distribution of the scaled energy fluctuations $\delta
E / \langle E \rangle_{hcs}$ . An example, corresponding to
$\alpha=0.8$ and $n_{H}=0.02 \sigma^{-2}$, is shown in Fig.
\ref{f4}. The results are well fitted by a Gaussian, within the
numerical precision of the data. In particular, no asymmetry is
observed. Up to date, there is no theoretical explanation
available for this simple behavior in an inherent non-equilibrium
system.

\section{Discussion}

In this paper, several questions related with the description of
dilute granular gases have been addressed. The basic theoretical
ingredient for the discussion has been the identification of the
hydrodynamic part of the spectrum of the linearized Boltzmann
equation for smooth inelastic hard spheres or disks. In the limit
of vanishing wavevector, the hydrodynamic modes are determined in
a unique way by the exact balance equations for the density,
momentum, and energy, for all values of the coefficient of
restitution. Beyond this limit, one has to resort to some
approximations, for instance perturbation expansions, assuming
analyticity and convergence. Moreover, if the hydrodynamic modes
dominate for long times and long wavelenghts, the hydrodynamic
Navier-Stokes equations for a dilute granular gas follow, as
discussed in Sec. \ref{s3}. Therefore, the fundamental and
controversial question of the existence of a hydrodynamic
description for granular gases described by the Boltzmann equation
can be sharply formulated as: are the hydrodynamic eigenvalues
smaller than all the other parts of the spectrum? For elastic
systems, this question has been positively answered \cite{McL89},
and it can be expected that much more be learned about the
spectrum of the inelastic linear Boltzmann equation in the near
future.

The procedure indicated above leads to formal expressions for the
transport coefficients of a dilute granular gas that can be
expressed in the form of Green-Kubo relations. The energy
dissipation in collisions introduces significant differences as
compared with the usual relations for molecular, elastic gases.
The expressions  have been evaluated by means of an $N$-particle
simulation technique. The results are in quite good agreement with
the analytical expressions obtained in the first Sonine
approximation for not too strongly inelastic systems, but
significant discrepancies are observed for small values of the
coefficient of restitution. An analysis of the simulation data
show that they are, at least in part, due to the presence of
relevant velocity correlations in the HCS, even in the low density
limit. At what extension this implies that the effect of  velocity
correlations should be incorporated somehow in the theoretical
analysis of the transport coefficients of a dilute granular gas,
is an open question. A related question is the role of velocity
correlations for finite density. It has been found that velocity
correlations might become important quite before a granular system
develops significant short range spacial correlations
\cite{ByR04}. Therefore, extensions of the Boltzmann equation to
finite density should include the effect of velocity correlations
in the precollisonal sphere. This offers another challenge to the
researchers on granular fluids.

As a necessary first step in the analysis of the velocity
correlations in the HCS of granular gases, a distribution function
describing partially their presence has been computed. Again,
advantage has been taken of the knowledge of the hydrodynamic
modes for vanishing wavevector. The identified velocity
correlations have an intrinsic inelastic character, vanishing in
the elastic limit, although the theoretical analysis carried out
predicts that they stay small for all values of the restitution
coefficient. The presence of velocity correlations implies that
the total energy of the system fluctuates. We have compared the
theoretical predictions for these fluctuations with molecular
dynamics results and found a good agreement. Nevertheless, due to
limitations associated with the instability of the HCS, the
simulation results are restricted to not very strong inelasticity.
The issue of whether  the theory also works in the very small
$\alpha$ region deserves future work.

\ack

This research was supported in part by the Ministerio de Ciencia y
Tecnolog\'{\i}a (Spain) through Grant No. BFM2002-00307 (partially
financed by FEDER funds).

\end{document}